\documentclass[%
 aip,
cha,
 amsmath,amssymb,
preprint,
]{revtex4-2}

\usepackage{graphicx}
\usepackage{dcolumn}
\usepackage{bm}

\usepackage[utf8]{inputenc}
\usepackage[T1]{fontenc}
\usepackage{mathptmx}
\usepackage{etoolbox}

\begin{document}


\title{On a hyperbolic Duffing oscillator with linear damping and periodic forcing} 



\author{Alain M. Dikand\'e}
\email{dikande.alain@ubuea.cm}
\affiliation{Laboratory of Research on Advanced Materials and Nonlinear Science, Department of Physics, Faculty of Science, University of Buea PO Box 63, Buea, Cameroon}

\date{\today}

\begin{abstract}
The Duffing oscillator is a textbook model that describes the dynamics of a point mass suspended on a spring with position-dependent stiffness. The point mass is assumed to experience a linear damping as well as a time-dependent periodic external forcing. The model has been instrumental in theoretical investigations of dynamical properties of systems with parity-conserving symmetry, where a double-well substrate connects two metastable states separated by a barrier. Physical systems of interest include nonlinear feedback-controlled mass-spring-damper oscillators, active hysteresis circuits (e.g. memristors), protein chains prone to hydrogen bond-mediated conformational transitions, centro-symmetric crystals and so on. In this work we consider a Duffing-type oscillator with a double-well potential represented by a hyperbolic function of mass position. The hyperbolic double-well potential has two degenerate minima that can be smoothly tuned by varying a deformability parameter, leaving unchanged the barrier height. We investigate solutions of the equation of motion in the absence and presence of damping and forcing. In the absence of perturbations numerical solutions lead to a periodic train of anharmonic oscillations featuring a crystal of pulse solitons of sech types. However, when the hyperbolic double-well potential is inverted, analytical solutions can be obtained which turn out to be kink-soliton crystals described by Jacobi elliptic functions. When damping and forcing are taken into consideration, the system dynamics can transit from periodic to chaotic phases or vice-versa via period-doubling or period-halving bifurcations, by simply varying the deformability parameter. The Poincar\'e map of the proposed model carries the well-known characteristic signatures of chaos presursors of the standard Duffing model, which happens to be just a particular case of the bistable oscillator model with the hyperbolic double-well potential.

\end{abstract}

\maketitle 

\section{\label{sec1} Introduction}
Double-well models (see e.g. \cite{a1,a2,a3,a4,a5,a51,a6}) are ubiquitous in physical systems where the symmetry is dominated by bistability, a feature associated with an inversion symmetry in structural properties of the systems. In these specific physical contexts the collective organization of constituants which can be atoms, molecules, ions, elementary circuits or let us say oscillators to generalize, is dominated by spatial and temporal configurations in which each oscillator disposes of two degenerate (i.e. equipotential) metastable states separated by an energy barrier. Physical systems of such types abound in nature, they are found among solids such as perovskites and centrosymmetric crystals \cite{a1,a3}. In biophysics and biochemistry they are present in chemical reactions where hydrogen bridges link molecular groups of the same or different species \cite{a5,a6,a7,a8,a81}, and in chemical oscillations involving tunnelings through energy barriers \cite{a9,a10}. Double-well models have played a fundamental role in the modern pictures of structural transitions in solids, soft matters including polymers \cite{a9,a10,a12,a13,a14}, and biological systems where conformational changes occur in close analogy with second-order transitions observed in perovskites and centro-symmetric crystals in general. In biomolecular systems such as long protein chains for instance \cite{a9,a10}, the double-well profile of the conformational energy suggests an equipotential feature of decaying and nascent protein conformations.\\
The Duffing-oscillator model \cite{a15} has been extremely insightful in theoretical investigations of dynamical properties of double-well systems, particularly in the physical contexts where damping and external forcing play significant roles in the system dynamics. The double-well substrate in this model is the $\phi^4$ potential \cite{a1,a2,a3}, which describes a bistable system with a configurational energy of a well fixed double-well profile. Namely the positions of its two degenerate minima are always fixed as are the height of the potential barrier, the stiffness of potential walls and the confinement of potential wells. Such a fixed shape can in some physical contexts turn out to be a drawback because it confers the $\phi^4$ model with a universal character, such that it can only address qualitative aspects of dynamical properties of concerned systems. Yet nature abounds with a very large variety of bistable systems exhibiting highly complex configurations, characterized by structures in which the double-well feature of the configurational energy is determined by the oscillators' environment. \\
Apart from the double-Morse potential \cite{a16} which is probably the most popularly known double-well substrate, some other double-well susbtrates with deformable shape profiles have been proposed \cite{a17,a18,a19,a20}. In recent works \cite{a21,a22,a23} a family of hyperbolic potentials, whose double-well shapes could be tuned distinctly by varying a single deformability parameter, was introduced. In one of them, positions of the two potential minima were shown to be tunable continuously leaving unaffected the height of the potential barrier. Physical systems displaying such features could be colonies of bacteries interacting through chemosensors and oscillating between two metastable states, glycolitic suspensions
of yeast cells oscillating in unison between two equilibrium states \cite{a24}, a slender steel suspended on a periodically driven rigid insulating frame whose bottom end hands in between two permanent magnets, or an active circuit with hysteretic nonlinear feedback. In these specific systems the environements of oscillators impact their equilibrium states. \\
In this work we consider a Duffing-type model in which the restoring force is provided by a hyperbolic double-well potential. This hyperbolic double-well potential is particular in that positions of its two stable equilibrium states can be tuned by varying a parameter, without affecting the barrier height. A slender steel trapped between two permanent magnets will perform excursions between two metastable states determined by positions of the two magnets. Besides this interesting mechanical setup we may also mention several other similar physical systems. We first look for solutions of the equation of motion in the absence of damping and forcing, considering two different physical contexts of hyperbolic potential namely the case when the potential is effectively a double-well field, and the case when it is a double-hump field. Then we examine the regime of damped and periodically forced motion by exploring time series and phase portraits of the system, generated numerically for selected values of the deformability parameter. While the standard Duffing model is known to exhibit a cascade of supercritical pitchfork bifurcations with respect to the magnitude of the external force, it will emerge that the deformability parameter also can lead the system dynamics through stable and unstable regimes or vice-versa via period-doubling or period-halving (i.e. supercritical and subcritical pitchfork) instabilities. The texture of Poincar\'e maps of the hyperbolic double-well oscillator will attest of common features with the standard Duffing model, making the hyperbolic double-well map an ideal generalization of the Duffing model.

\section{\label{sec2} The model and fixed-point solutions}
\begin{figure}\centering
\includegraphics[width=4.in, height= 2.5in]{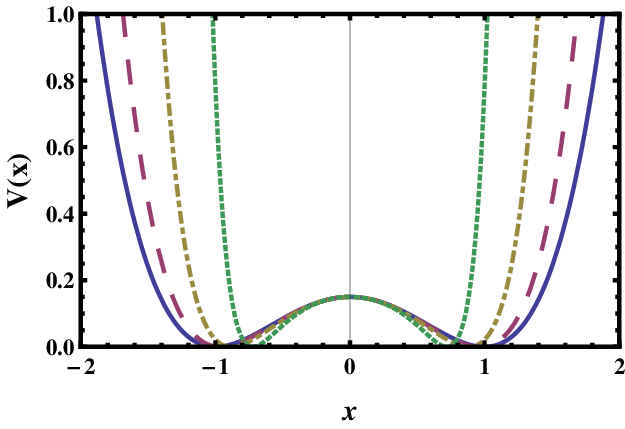}
\caption{The hyperbolic double-well potential $V_{\mu}(x)$ given by (\ref{eq2}) versus position $x$, for four different values of $\mu$ namely $\mu=0.1$ (solid curve), $\mu=0.5$ (long dashed curve), $\mu=1.$ (dot-dashed curve) and $\mu=2$ (dashed curve).}
\label{fig1}
\end{figure}

Consider an oscillator of finite mass $m$, moving in the field of force generated by a position-dependent anharmonic potential $V_{\mu}(x)$. In the presence of linear damping and time-varying periodic forcing, the equation of motion of the oscillator can be written:
\begin{equation}
 \ddot{x} + \gamma\dot{x}= - \frac{1}{m}\, \frac{\partial V_{\mu}(x)}{\partial x} + f_0\cos(\omega t). \label{eq1}
\end{equation}
$\gamma$ and $f_0$ are respectively the coefficient of linear damping and the magnitude of the driving force, all relative to the oscilator mass $m$. Without loss of generality we shall set $m\equiv 1$. \\
The quantity $V_{\mu}(x)$ in eq. (\ref{eq1}) is the substrate potential, assumed to depend on the mass position $x$ through the following hyperbolic function:
 \begin{equation}
V_{\mu}(x)=\frac{a_0}{4}\biggl(\frac{1}{\mu^2}\sinh^2\left(\mu\, x\right) - 1 \biggr)^2, \qquad \mu\neq 0,\label{eq2}
 \end{equation}
with $a_0$ and $\mu$ two real parameters. $a_0$ is proportional to the amplitude of the potential barrier and is positive for bistable systems with double-degenerate minima, and negative when the system dynamics takes place in an anharmonic potential with two energy humps (two degenerate maxima) surrounding an energy well (minimum). Although $\mu$ is required to be nonzero, it can be shown that when $\mu\rightarrow 0$ the hyperbolic double-well potential (\ref{eq2}) reduces to the $\phi^4$;
\begin{equation}
V(x)=\frac{a_0}{4}\biggl(x^2 - 1\biggr)^2. \label{f4}
\end{equation}
 The graph of fig.\ref{fig1} depicts the hyperbolic potential eq. (\ref{eq2}), for positive $a_0$ and for four different values of the deformability parameter $\mu$.
\par Fixed points of eq. (\ref{eq1}) are its equilibrium solutions in the absence of damping and external forcing. So to say they are just the extrema of the double-well potential $V_{\mu}(x)$. for $a_0>0$ there are three extrema, namely $x_0=0$ which is a maximum where a potential wall is erected, and two degenerate minima at $x_{1,2}$ with:
 \begin{equation}
x_{1,2}=\pm \frac{arcsinh(\mu)}{\mu}.
 \end{equation}
The height of the potential barrier, $V(0)=a_0/4$, clearly remains insensitive to change of the deformability parameter $\mu$ as seen in fig. \ref{fig1}.

\section{\label{sec3} Nonlinear solutions in the regime of free motion}
\begin{figure}\centering
\begin{minipage}[c]{0.34\textwidth}
\includegraphics[width=2.1in, height= 1.5in]{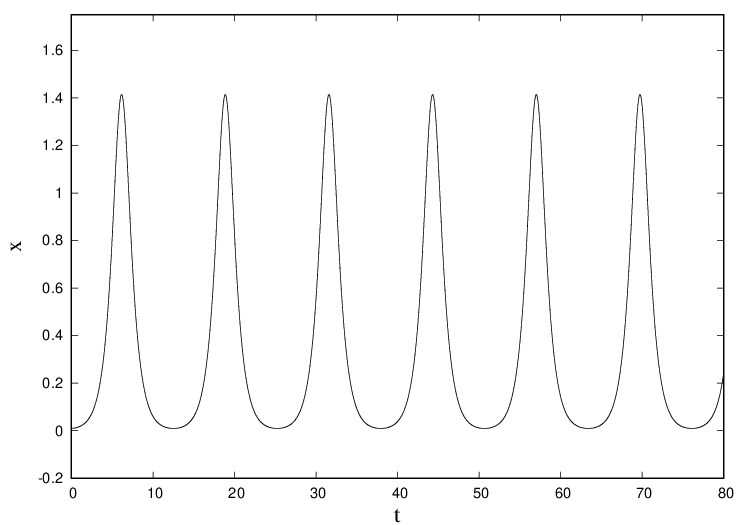}
\end{minipage}%
\begin{minipage}[c]{0.34\textwidth}
\includegraphics[width=2.1in, height= 1.5in]{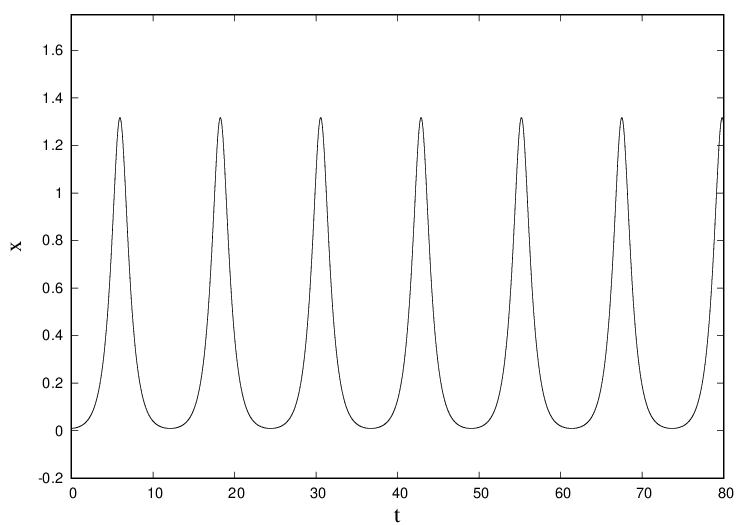}
\end{minipage}%
\begin{minipage}[c]{0.34\textwidth}
\includegraphics[width=2.1in, height= 1.5in]{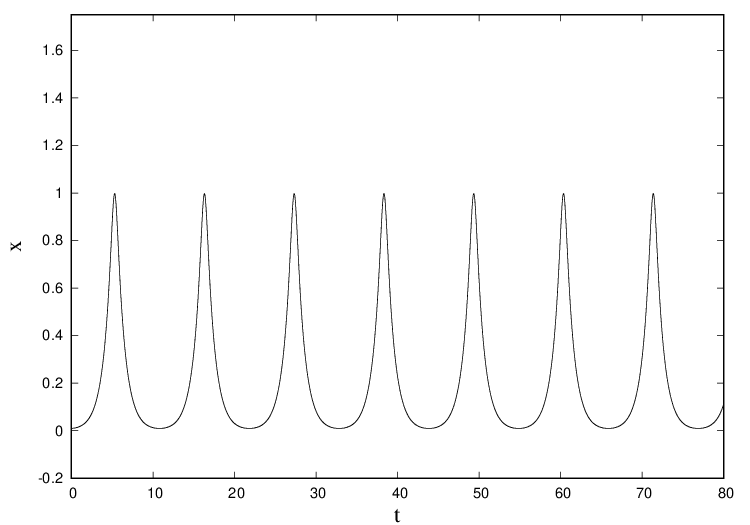}
\end{minipage}
\caption{Time trace $x(t)$ in the unbiased regime for $\mu=0.0005$ (left graph), $\mu=0.5$ (middle graph) and $\mu=1.5$ (right graph).}
\label{fig2}
\end{figure}

\begin{figure}\centering
\begin{minipage}[c]{0.51\textwidth}
\includegraphics[width=3.in, height= 2.5in]{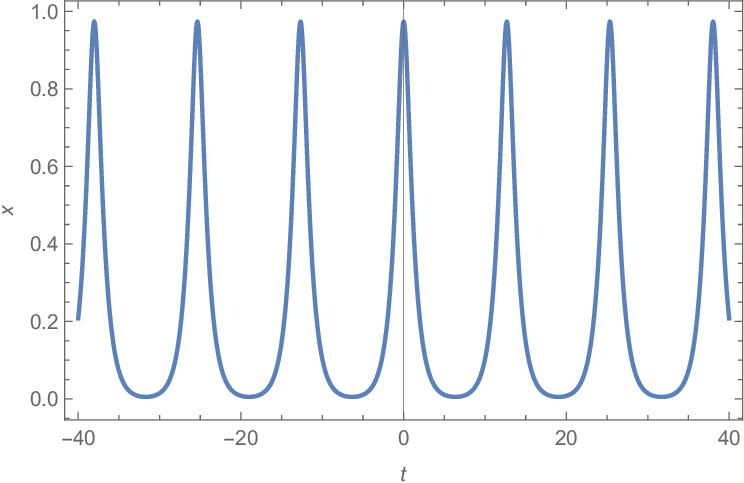}
\end{minipage}%
\begin{minipage}[c]{0.51\textwidth}
\includegraphics[width=3.in, height= 2.5in]{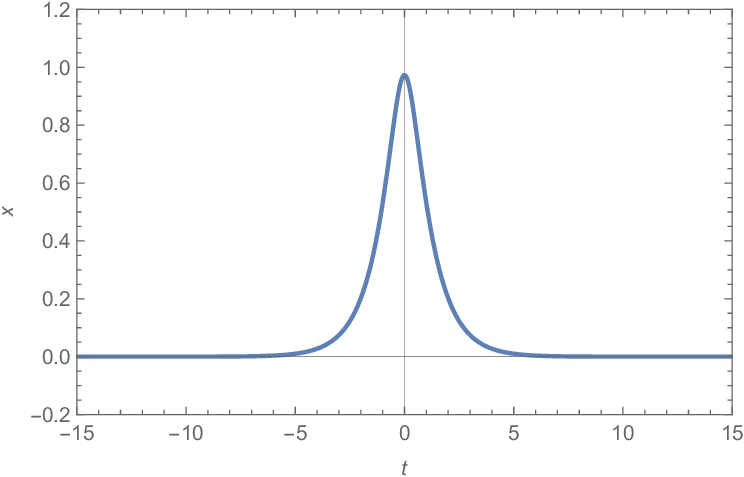}
\end{minipage}
\caption{Plot of the hypothetical analytical solution (\ref{eq7}), of eq. (\ref{eq1}) when $a_0>0$.}
\label{fig3}
\end{figure}

In the absence of damping and forcing, equation (\ref{eq1}) becomes:
\begin{equation}
 \ddot{x} + \frac{\partial V_{\mu}(x)}{\partial x} = 0. \label{eq3}
\end{equation}
To find solutions to eq. (\ref{eq3}) it is convenient to express the latter in the quadrature form i.e.:
\begin{equation}
 E=\frac{\dot{x}^2}{2} + V_{\mu}(x), \label{eq4}
\end{equation}
where the integration constant $E$ can readily be identified as the conserved energy. Equation (\ref{eq4}) thus leads to the following expression for the oscillator velocity:
\begin{equation}
\dot{x}= \sqrt{2\biggl(E-V_{\mu}(x)\biggr)}. \label{vit}
\end{equation}
Equation (\ref{vit}) is relevant in our quest of solutions of equation (\ref{eq3}), for it suggests that both the sign of $E$ and its magnitude relative to $a_0$ will matter. Thus, if $E$ is real and positive as expected a real solution will be possible only when $E$ is above the potential barrier. Note that we may also have real solutions for negative $a_0$, interesting enough this second case offers a possibility for an analytical solution when $E=0$. But before considering this second possibility let us first find solutions to equation (\ref{eq4}) for positive $a_0$. Proceeding with the study we solve equation (\ref{eq4}) numerically with the help of a sixth-order Runge-Kutta algorithm with fixed step \cite{luth}. Fig. \ref{fig2} displays the numerical solution $x(t)$ with the three graphs corresponding to three different values of $\mu$, which are $\mu=0.0005$ (left graph), $\mu=0.5$ (middle graph) and $\mu=1.5$ (right graph).
The temporal profile of $x(t)$ in fig. \ref{fig2} is reminiscent of a periodic train of sech-type pulse solitons. To check the consistency of our conjecture let us pick the following analytical function, which we assume to be a suitable analytical solution equivalent to the numerical solution in fig. \ref{fig2}:
\begin{equation}
x(t)=x(0)\,arctanh\biggl[f(\mu,\kappa)\,dn \biggl(\frac{t}{\tau_{\kappa(E)}}\biggr)\biggr]. \label{eq7}
\end{equation}
In the above formula $dn(\cdots)$ is the elliptic Jacobi $dn$ function of modulus $\kappa(E)$ \cite{abra}. The hypothetical analytical solution (\ref{eq7}) is plotted in fig. \ref{fig3} versus $t$, for $\kappa=0.95$ (left graph) and $\kappa=1$ (right graph) ($x(0)$, $f(\mu,\kappa)$ and $\tau_{\kappa(E)}$ were assigned arbitrary values given that they are parameters that are time independent, as a matter of fact our assumption is correct so long as we are interested only in qualitative features of the analytical function). One see that figs. \ref{fig2} and \ref{fig3} (left graph) are qualitatively simailar, and so the solution is indeed a periodic train of sech-type pulse solitons. It is relevant to stress that the localized structure seen in the right graph of fig. \ref{fig3} is a single-pulse soliton of sech type, consistent with the fact that $dn(\cdots)\rightarrow sech(\cdots)$ when $\kappa(E)\rightarrow 1$. \cite{solduf1,solduf2,solduf3,solduf4}.

\begin{figure}\centering
\begin{minipage}[c]{0.51\textwidth}
\includegraphics[width=3.in, height= 2.5in]{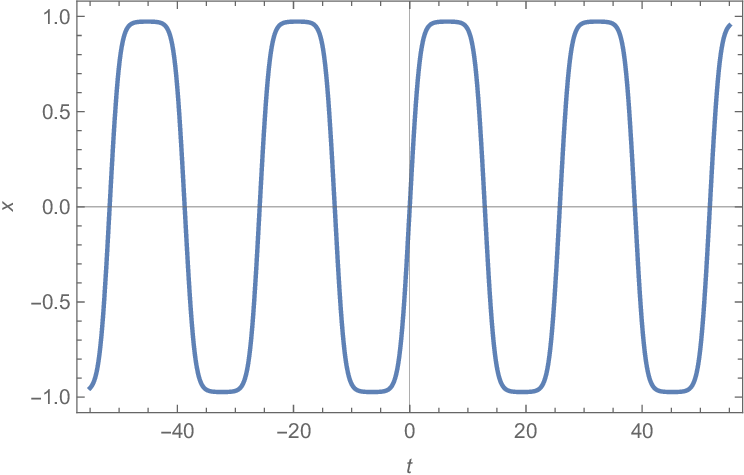}
\end{minipage}%
\begin{minipage}[c]{0.51\textwidth}
\includegraphics[width=3.in, height= 2.5in]{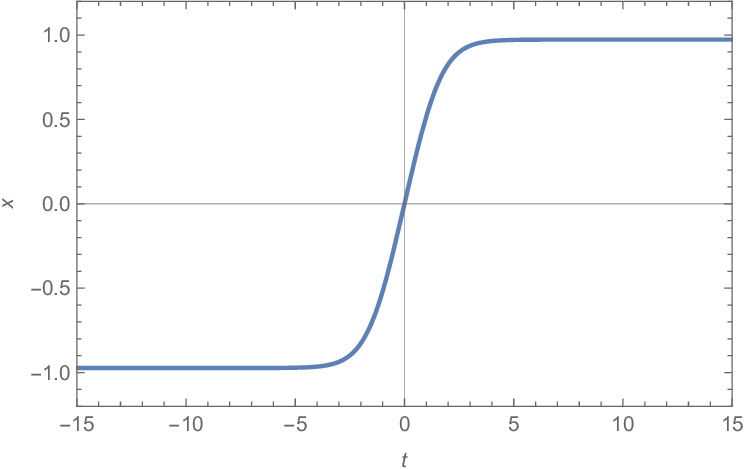}
\end{minipage}
\caption{Temporal profile of the nonlinear periodic solution $x(t)$ given by equation (\ref{eq6}), for $\kappa=0.98$ (left panel) and $\kappa=1$ (right panel). Remark that the right graph reproduces the single-kink soliton solution eq. (\ref{eq5}).}
\label{fig4}
\end{figure}

We argued that when $a<0$, it was possible to find an analytical solution to the equation of motion (\ref{eq1}). From the standpoint of mathematical physics this case is relevant and has indeed been addressed severaly in the context of $\phi^4$-based Duffing model \cite{solduf2,solduf4,wig,chao}. For $a_0<0$ the two symmetric wells of the bistable potential gets inverted and the two fixed points $x_{1,2}$ now turn to two degenerate maxima, while $x_0$ becomes a finite potential minimum. Let us seek analytical solutions to eq. (\ref{vit}) with negative $a_0$, for two distinct boundary conditions: non-periodic boundary condition ($0 \leq t \leq \infty$), and periodic boundary condition ($-\tau_0 \leq t \leq \tau_0$). The first boundary condition amounts to looking for solution that behaves like $x\rightarrow x_{1,2}$ as $t\rightarrow \pm \infty$, implying $E=0$. Integrating the velocity equation (\ref{vit}) exactly we obtain the following analytical solution:
\begin{equation}
x(t)=\frac{1}{\mu}\,arctanh\biggl[\frac{\mu}{\sqrt{1+\mu^2}}\,\tanh\biggl(\frac{t}{\tau}\biggr)\biggr], \qquad \tau=\sqrt{\frac{-2}{a_0(1+\mu^2)}} \label{eq5}
\end{equation}
Eq. (\ref{eq5}) is a single-kink soliton of half tail $1/\sqrt{1+\mu^2}$ and average width $\tau$. As for the periodic boundary condition, we require a solution behaving like  $x(t\pm\tau_0)=x(t)$ where $\tau_0$ is a finite time. With this requirement an integration of equation (\ref{vit}) yields:
\begin{eqnarray}
x(t)&=&\frac{1}{\mu}\,arctanh\biggl[\frac{\mu\,\kappa(E)}{\sqrt{1+\mu^2}}\,\sqrt{\frac{2}{1+\kappa^2(E)}}\,sn \biggl(\frac{t}{\tau_{\kappa(E)}}\biggr)\biggr], \nonumber \\
\tau_{\kappa(E)}&=&\sqrt{\frac{-(1+\kappa^2(E))}{a_0(1+\mu^2)}}, \label{eq6}
\end{eqnarray}
where $sn$ is the elliptic Jacobi $sn$ function \cite{abra} of modulus $\kappa(E)$. Instructively $\kappa(E)$ is determined by $E$, and here also we still have the requirement $0\leq \kappa(E) \leq 1$. When $\kappa(0)=1$ the Jacobi elliptic function $sn(\cdots)=\tanh(\cdots)$ and eqs. (\ref{eq6}) reduces exactly to the single-kink solution eq. (\ref{eq5}). Fig. \ref{fig4} shows plot of the nonlinear periodic solutions (\ref{eq6}) for $\kappa(E)=0.98$ (left graph) and $\kappa(0)=1$ (right graph), considering only one value of $\mu$ (i.e. $\mu=0.1$) given that the $\mu$ dependences of characteristic parameters of the single-kink soliton, particularly its width $\tau_{\kappa(E)}$, can be assessed from its analytical expression given in formula (\ref{eq6}).
\section{\label{sec4} Dynamics in the presence of damping and periodic forcing}
\begin{figure}\centering
\begin{minipage}[c]{0.51\textwidth}
\includegraphics[width=3.in, height= 2.in]{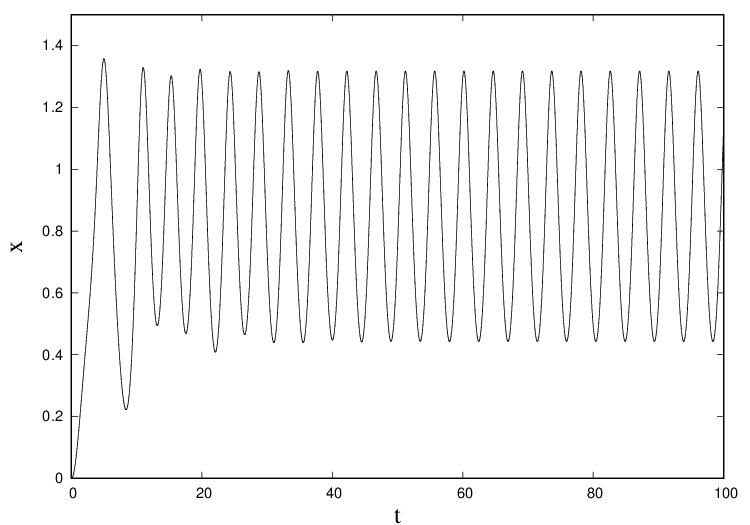}
\end{minipage}%
\begin{minipage}[c]{0.51\textwidth}
\includegraphics[width=3.in, height= 2.in]{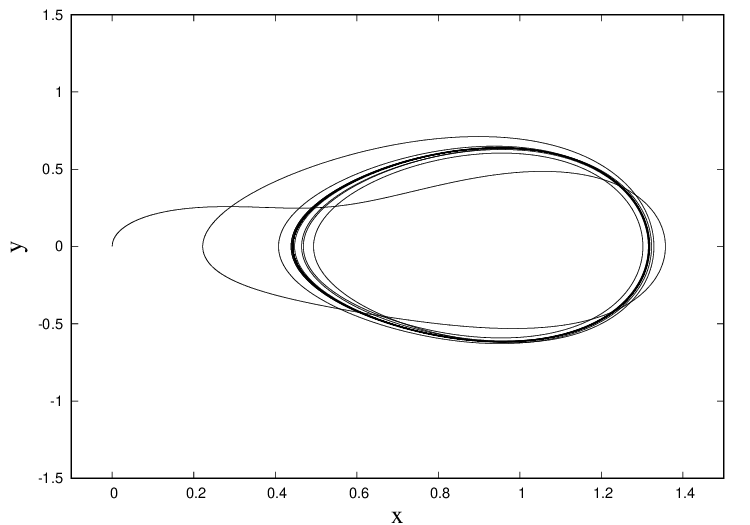}
\end{minipage}\\
\begin{minipage}[c]{0.51\textwidth}
\includegraphics[width=3.in, height= 2.in]{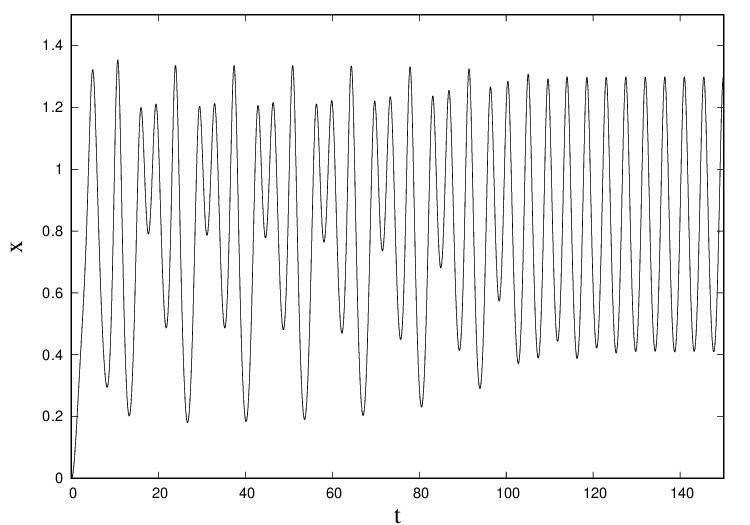}
\end{minipage}%
\begin{minipage}[c]{0.51\textwidth}
\includegraphics[width=3.in, height= 2.in]{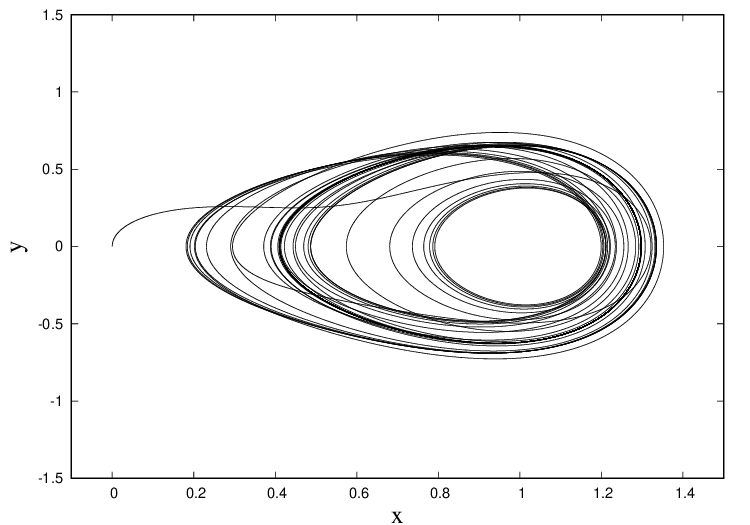}
\end{minipage}\\
\begin{minipage}[c]{0.51\textwidth}
\includegraphics[width=3.in, height= 2.in]{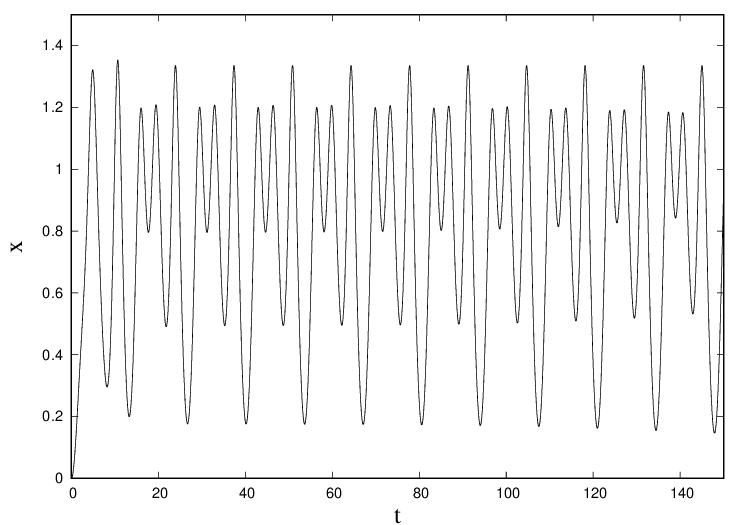}
\end{minipage}%
\begin{minipage}[c]{0.51\textwidth}
\includegraphics[width=3.in, height= 2.in]{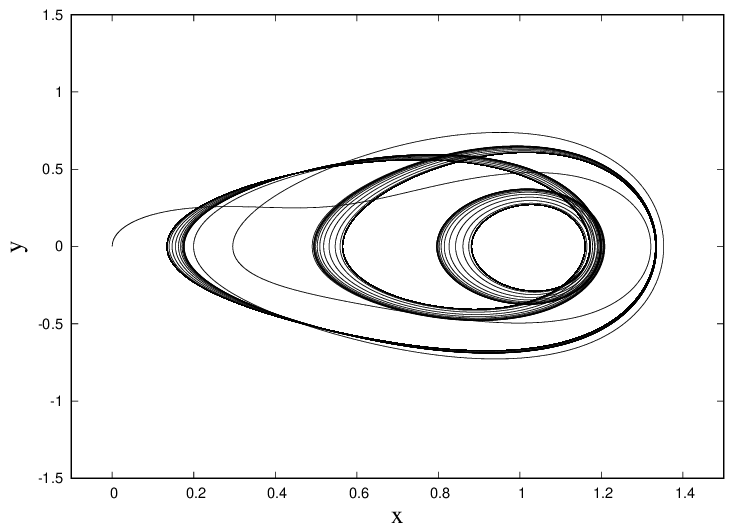}
\end{minipage}\\
\begin{minipage}[c]{0.51\textwidth}
\includegraphics[width=3.in, height= 2.in]{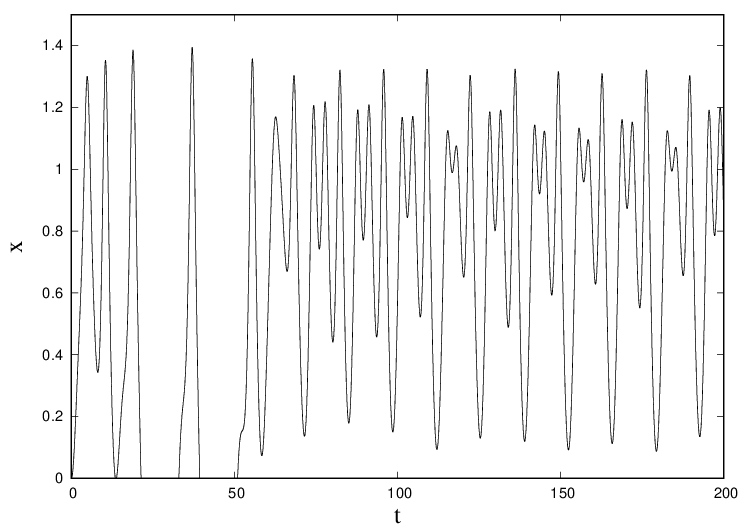}
\end{minipage}%
\begin{minipage}[c]{0.51\textwidth}
\includegraphics[width=3.in, height= 2.in]{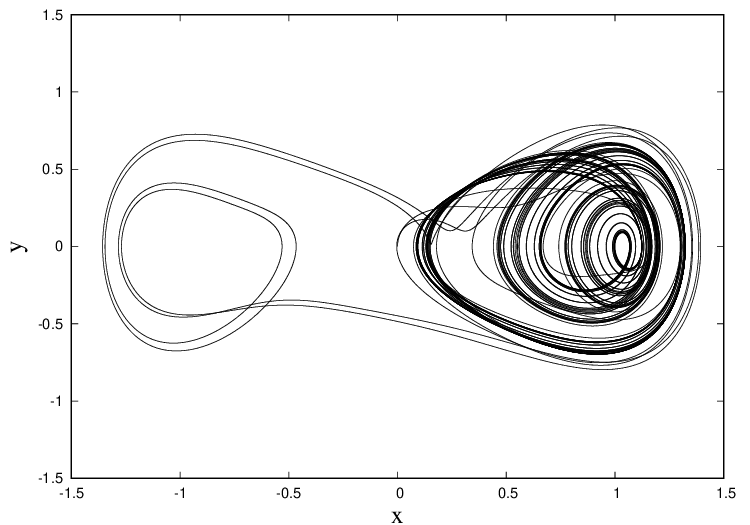}
\end{minipage}
\caption{Time series (left graphs) and phase portraits (right graphs) of the hyperbolic double-well map for $f_0=0.3$, $\gamma=0.25$ and four different values of $\mu$. From top to bottom row different graphs correspond to $\mu=10^{-6}$ (Duffing model), $\mu=0.271$, $\mu=0.272$, $\mu=0.35$.}
\label{fig5a}
\end{figure}

\begin{figure}\centering
\begin{minipage}[c]{0.51\textwidth}
\includegraphics[width=3.in, height= 2.in]{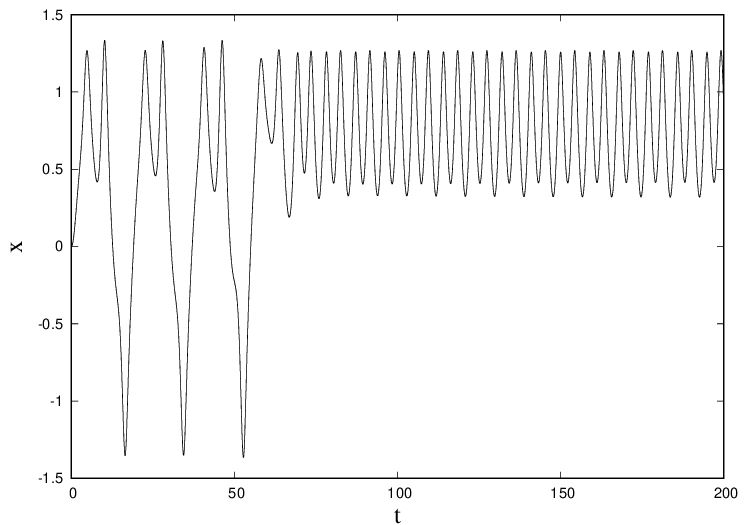}
\end{minipage}%
\begin{minipage}[c]{0.51\textwidth}
\includegraphics[width=3.in, height= 2.in]{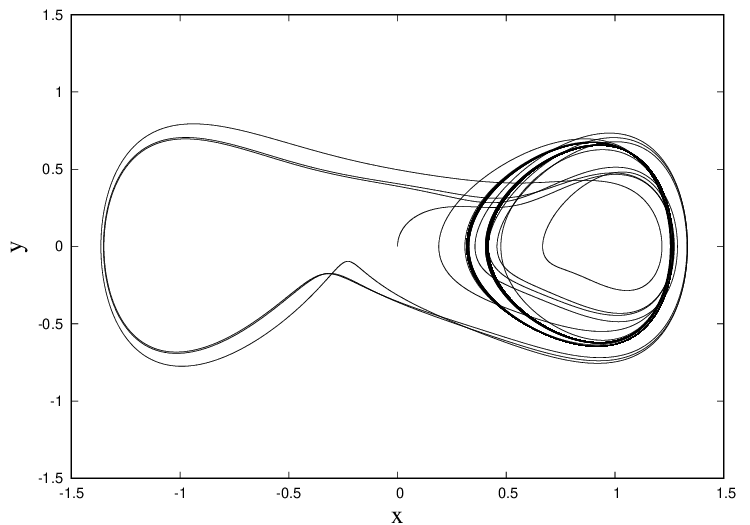}
\end{minipage}\\
\begin{minipage}[c]{0.51\textwidth}
\includegraphics[width=3.in, height= 2.in]{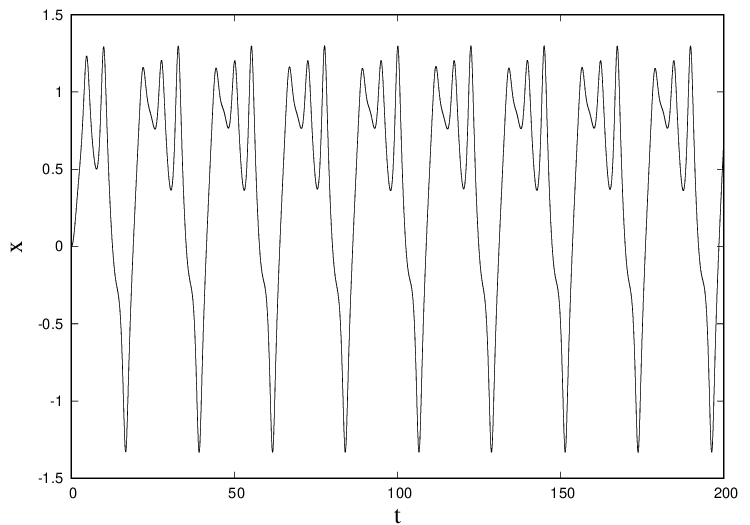}
\end{minipage}%
\begin{minipage}[c]{0.51\textwidth}
\includegraphics[width=3.in, height= 2.in]{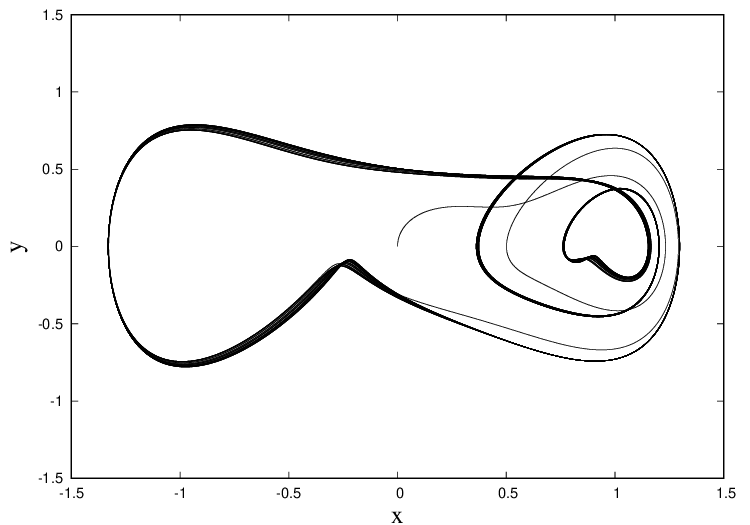}
\end{minipage}\\
\begin{minipage}[c]{0.51\textwidth}
\includegraphics[width=3.in, height= 2.in]{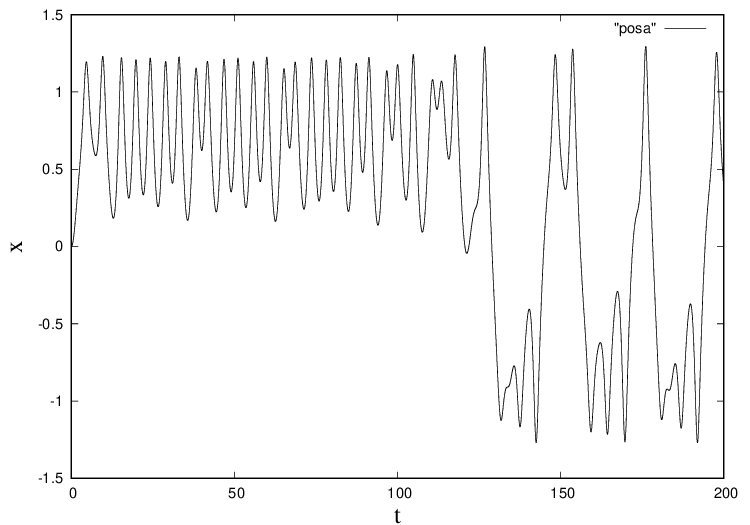}
\end{minipage}%
\begin{minipage}[c]{0.51\textwidth}
\includegraphics[width=3.in, height= 2.in]{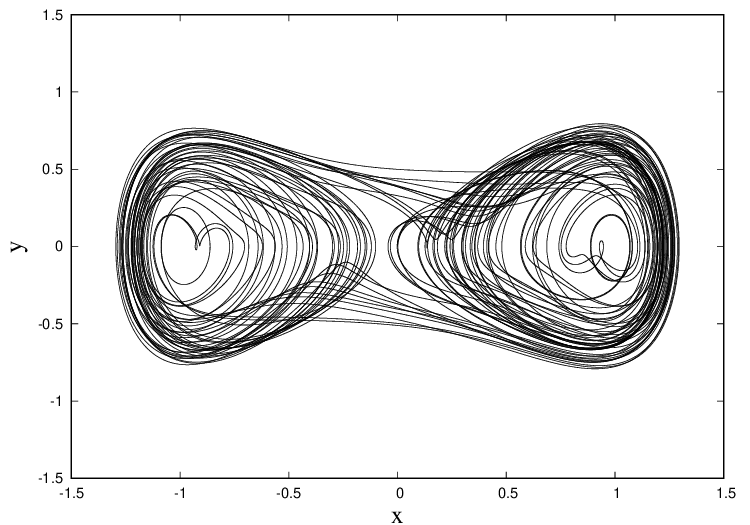}
\end{minipage}\\
\begin{minipage}[c]{0.51\textwidth}
\includegraphics[width=3.in, height= 2.in]{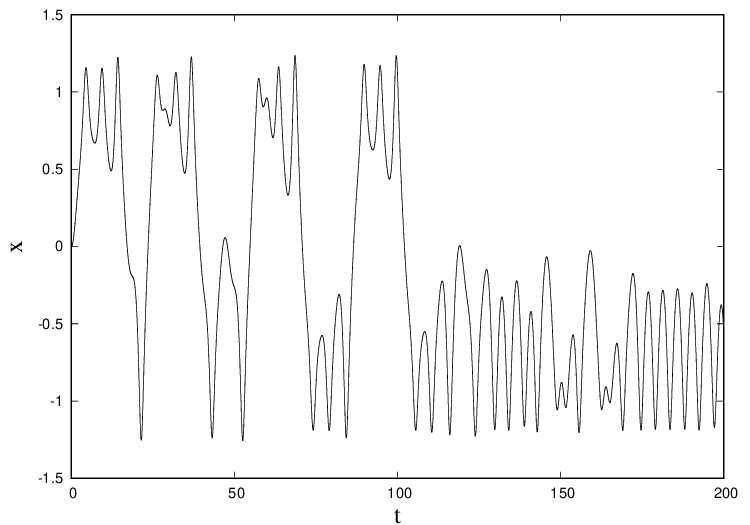}
\end{minipage}%
\begin{minipage}[c]{0.51\textwidth}
\includegraphics[width=3.in, height= 2.in]{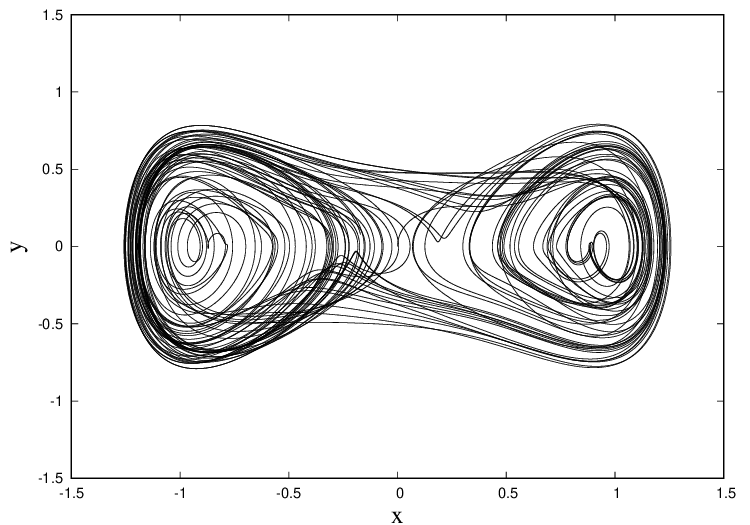}
\end{minipage}
\caption{Time series (left graphs) and phase portraits (right graphs) of the hyperbolic double-well map for $f_0=0.3$, $\gamma=0.25$ and four different values of $\mu$. From top to bottom rows different graphs correspond to $\mu=0.45$, $\mu=0.55$, $\mu=0.65$, $\mu=0.75$.}
\label{fig5b}
\end{figure}

\begin{figure}\centering
\includegraphics[width=6.in, height= 2.5in]{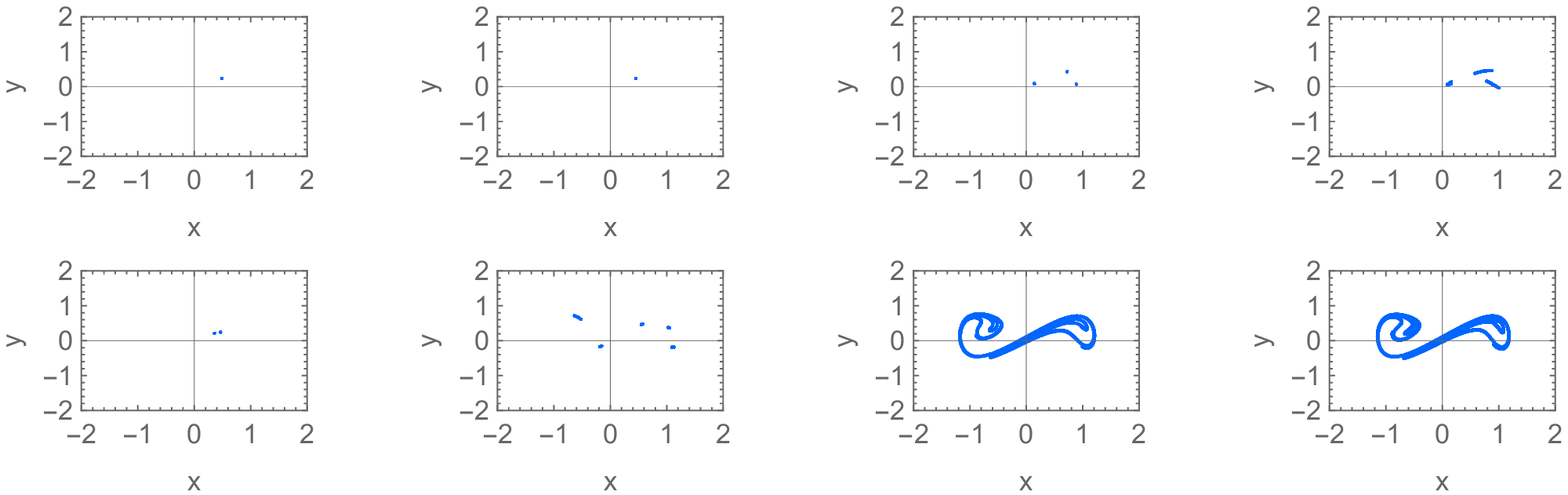}
\caption{Poincar\'e map of the two-dimensional hyperbolic double-well map for $f=0.3$, $\gamma=0.25$ and $\omega=1.4$. Top row, from left graph to right graph: $\mu=10^{-6}$, $\mu=0.271$, $\mu=0.272$, $\mu=0.35$. Bottom row, from left graph to right graph: $\mu=0.45$, $\mu=0.55$, $\mu=0.65$, $\mu=0.75$.}
\label{fig6}
\end{figure}

We now turn to the case when the motion of the oscillator is perturbed by a linear damping and an external periodic forcing, focusing mainly on the motion in a double-well potential (i.e. $a>0$). In this context it is useful to first stress that when $\mu \rightarrow 0$, eq. (\ref{eq2}) becomes eq. (\ref{f4}) such that the nonlinear damped and driven equation (\ref{eq1}) reduces to:
\begin{equation}
 \ddot{x} + \gamma\dot{x} - \omega_0^2\bigl(x -x^3\bigr) = f_0\cos(\omega t), \qquad  \omega_0^2=a_0. \label{eq8}
\end{equation}
The Duffing equation (\ref{eq8})\cite{solduf2} is a classic model of damped and forced bistable anharmonic oscillator with immensely rich dynamical properties, some of which can be summarized as follows: for $\gamma=f_0=0$, eq. (\ref{eq8}) possesses three extrema namely a saddle at $x=0$ which is an unstable fixed point, and two sinks at $x=\pm 1$ which are asymptotically stable fixed points. When damping and periodic forcing are switched on all trajectories of the oscillator will ultimately converge to a region enclosing the three equilibrium points \cite{wig}. Below some critical value of $\gamma$, determined by the magnitude $f_0$ and frequency $\omega$ of the external force \cite{wig}, the Poincar\'e map of the Duffing oscillator exhibits transverse homoclinic orbits to an hyperbolic fixed point and a profusion of periodic orbits embedded in chaotic attractors \cite{solduf2,wig}. \\ In view of the incommensurate richness of the standard Duffing model, it is clear that a single context of study such as the present one cannot enable us to unveil all of them. We will therefore be satisfied with only few key aspects of the dynamics of the damped and periodically driven hyperbolic double-well oscillator eq. (\ref{eq1}). To ease our analysis we reformulate the nonlinear damped and periodically forced equation (\ref{eq1}) as the two-dimensional nonlinear continuous map:
\begin{eqnarray}
\dot{x}&=& y, \nonumber \\
\dot{y}&=& \gamma y + F_{\mu}(x) + f_0\cos(\omega t), \qquad F_{\mu}=-\frac{\partial V_{\mu}(x)}{\partial x},  \label{eq9}
\end{eqnarray}
whose solutions are sought by numerical simulations of the related system of two coupled nonlinear ordinary differential equations. In the course of simulations we fixed values of some parameters, namely $\omega=1.4$ and $a_0=1$, whereas  the cofficient of linear dampling $\gamma$ and the magnitude of the external force $f_0$ will be given two different set of values (of course chosen on purpose, as it will be understood). Instructively the most interesting manifestations of regular and strange dynamical processes are expected for the oscillator's dynamics near its saddle point $(x, y)=(0, 0)$, where in the particular case of Duffing model trajectories are simple-periodic, multi-periodic, chaotic orbits or strange attractors embedded within limit cycles \cite{solduf2,toguy}.\\
We first consider numerical results of the two-dimensional continuous map (\ref{eq9}) for $f_0=0.3$ and $\gamma=0.25$. In fig. \ref{fig5a}, we plot the time series $x(t)$ (right graph) and phase portrait $y(x)$ (left graph) for $\mu=10^{-6}$, $\mu=0.271$, $\mu=0.272$ and $\mu=0.35$. In fig. \ref{fig5b} the same physical quantities are plotted for $\mu=0.45$, $\mu=0.55$, $\mu=0.65$ and $\mu=0.75$.
Graphs of the top row of fig. \ref{fig5a} suggest that when $\mu\rightarrow 0$, the system dynamics is dominated by simple-periodic anharmonic oscillations settling after a finite transient time. The trajectories however are not stricly periodic due to the damping and forcing that affect the oscillation amplitudes, making the oscillations quasi periodic for the parameter values chosen. It is to be recalled that for the value of $\mu$ corresponding to this specific behaviour, the two-dimensional continuous map is just the standard Duffing model. Now when $\mu$ increases, the quasi-periodic oscillations become more and more unstable in favor of multi-periodic structures. When $\mu$ is increased to a value above $0.271$ or a little higher, double-scroll orbits emerge with quasi-periodic or multi-periodic oscillations visiting the two sinks. The phase portrait of the bottom row of fig. \ref{fig5b} illustrates the emergence of a strange attractor quite characteristic of the standard Duffing model, yet the value of $\mu$ for this graph is relatively high i.e. $\mu=0.75$. Fig. \ref{fig6} summarizes rather well the evolution of characteristic features of the system dynamics when $\mu$ is varied, indeed the Poincar\'e maps in fig. \ref{fig6} generated for the eight values of $\mu$ exhibit the signature of period-doubling cascades as a result of the map parametrization.
\par
\begin{figure}\centering
\begin{minipage}[c]{0.51\textwidth}
\includegraphics[width=3.in, height= 2.in]{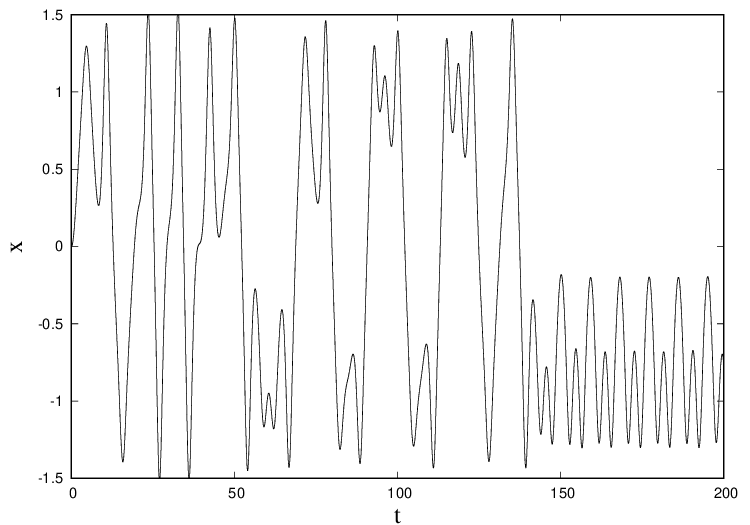}
\end{minipage}%
\begin{minipage}[c]{0.51\textwidth}
\includegraphics[width=3.in, height= 2.in]{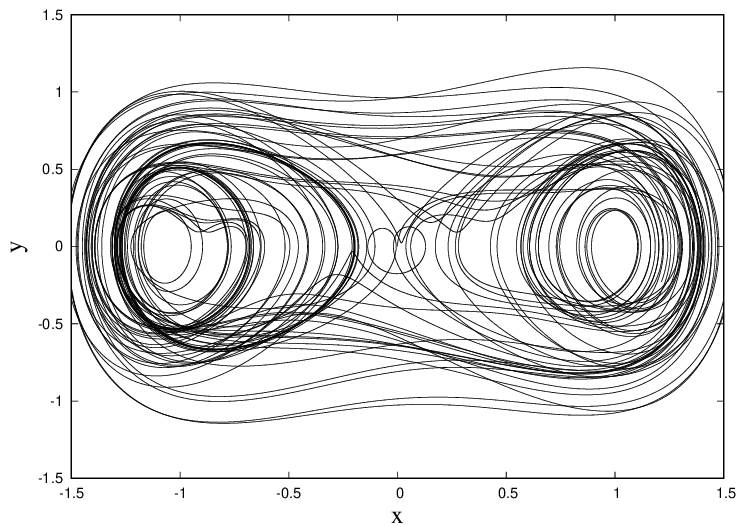}
\end{minipage}\\
\begin{minipage}[c]{0.51\textwidth}
\includegraphics[width=3.in, height= 2.in]{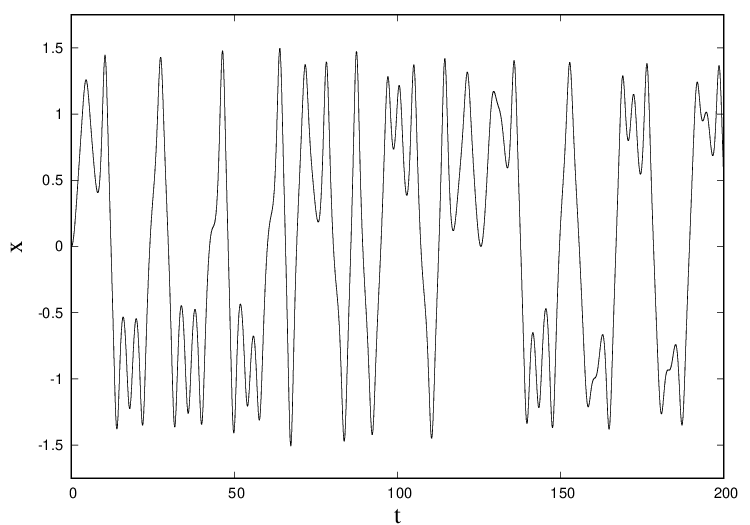}
\end{minipage}%
\begin{minipage}[c]{0.51\textwidth}
\includegraphics[width=3.in, height= 2.in]{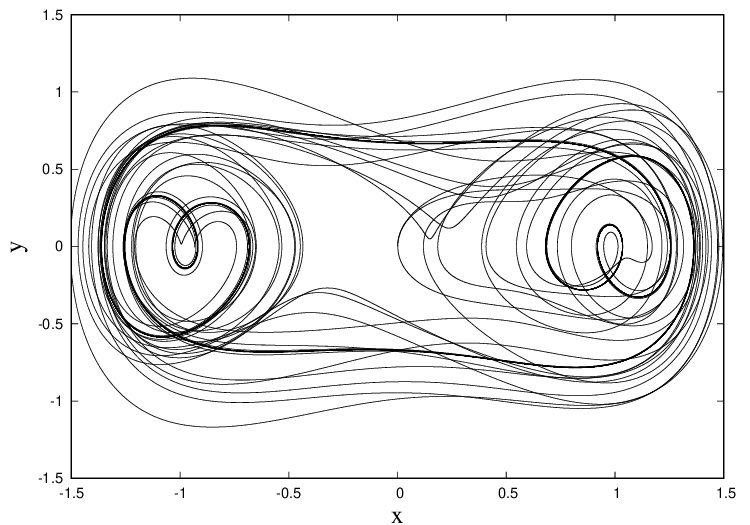}
\end{minipage}\\
\begin{minipage}[c]{0.51\textwidth}
\includegraphics[width=3.in, height= 2.in]{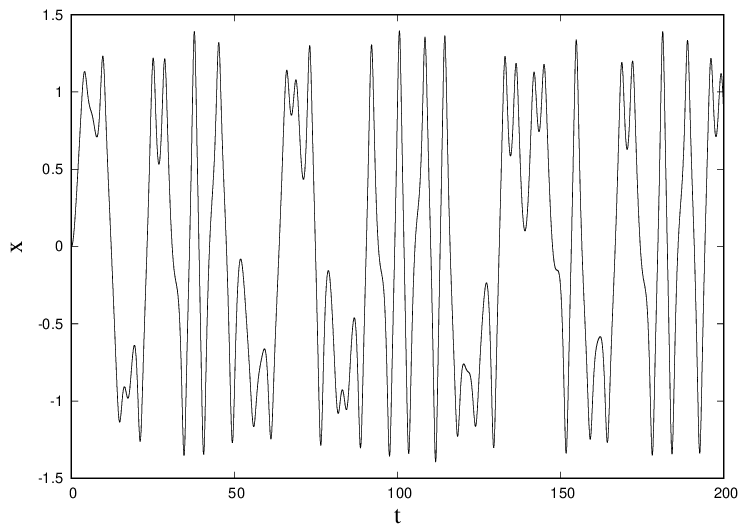}
\end{minipage}%
\begin{minipage}[c]{0.51\textwidth}
\includegraphics[width=3.in, height= 2.in]{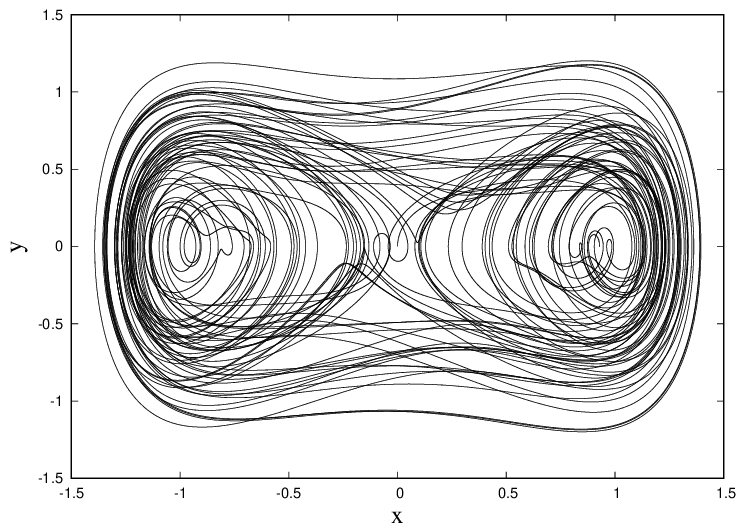}
\end{minipage}\\
\begin{minipage}[c]{0.51\textwidth}
\includegraphics[width=3.in, height= 2.in]{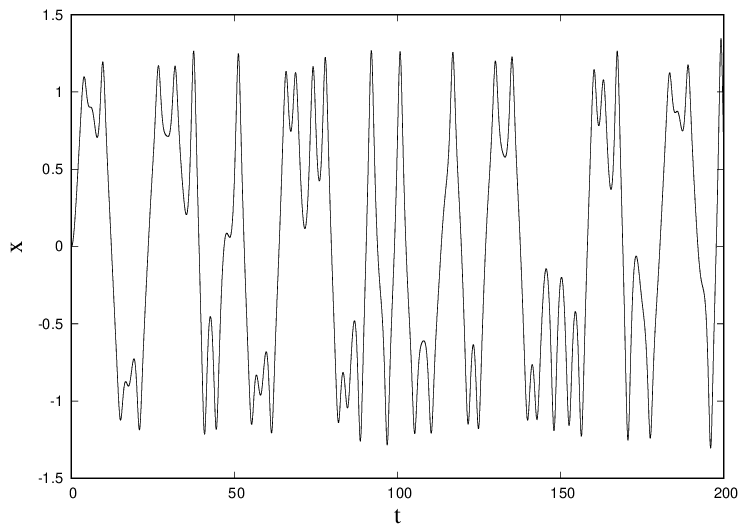}
\end{minipage}%
\begin{minipage}[c]{0.51\textwidth}
\includegraphics[width=3.in, height= 2.in]{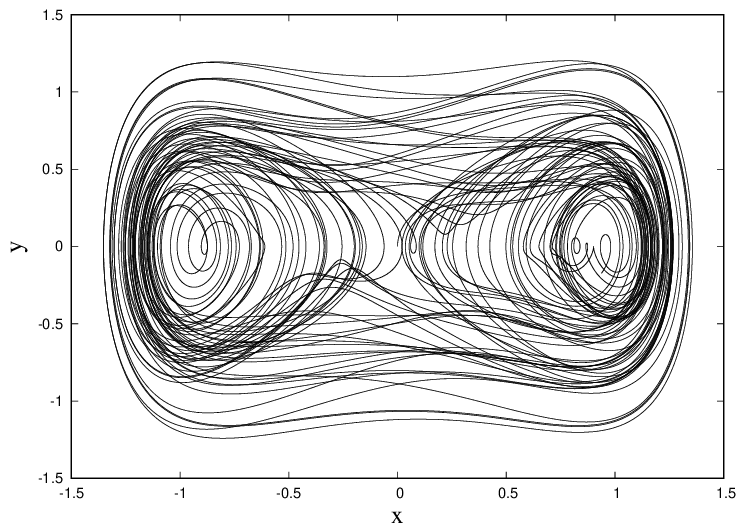}
\end{minipage}
\caption{Time series (left graphs) and phase portraits (right graphs) of the hyperbolic double-well map for $f_0=0.35$, $\gamma=0.1$, and four different values of $\mu$. From top to bottom rows different graphs correspond to $\mu=0.10^{-6}$ (Duffing model), $\mu=0.271$, $\mu=0.65$, $\mu=0.75$.}
\label{fig7}
\end{figure}

\begin{figure}\centering
\includegraphics[width=6.in, height= 2.5in]{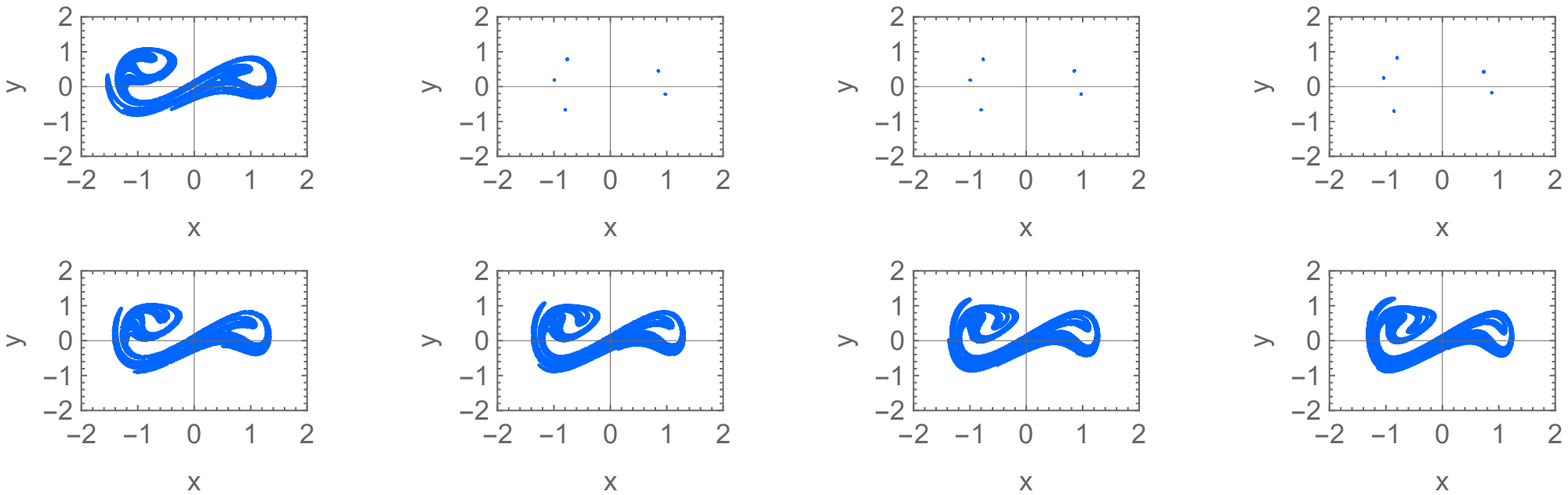}
\caption{Poincar\'e map of the two-dimensional hyperbolic double-well map for $f=0.35$, $\gamma=0.1$ and $\omega=1.4$. Top row, from left graph to right graph: $\mu=10^{-6}$, $\mu=0.271$, $\mu=0.272$, $\mu=0.35$. Bottom row, from left graph to right graph: $\mu=0.45$, $\mu=0.55$, $\mu=0.65$, $\mu=0.75$.}
\label{fig8}
\end{figure}

 What would be the trend in the evolution of dynamical features of the system, when $f_0$ and $\gamma$ are given values for which the standard Duffing oscillator is in the chaotic regime? It is well established that chaotic phenomena emerge in the standard Duffing oscillator when the magnitude of the external field $f_0$ is relatively higher compared to the damping coefficient $\gamma$. In Fig. \ref{fig7} we plotted the time series $x(t)$ and the phase portrait of the hyperbolic double-well oscillator for $f_0=0.35$ and $\gamma=0.1$, assuming four different values of the deformability parameter i.e. $\mu=10^{-6}$ (Duffing limit), $\mu=0.271$, $\mu=0.65$ and $\mu=0.75$. It is quite remarkable that for $\mu\rightarrow 0$, when the hyperbolic double-well map reduces to the standard Duffing model, the phase portait features coexisting strange attractors and a limit cycle.
The Poincar\'e map corresponding to this new set of parameter values (i.e. $f_0=0.35$ and $\gamma=0.1$), shown in fig. \ref{fig8}, suggests possible subcritical pitchfork transitions from chaotic to few-period oscillations for relatively high values of the deformability parameter. For instance we can clearly notice the emergence of five-period oscillations in the Poincar\'e maps for $\mu=0.271$, $\mu=0.272$ and $\mu=0.35$ succeeding a chaotic regime for $\mu=10^{-6}$, and then followed by chaotic attractors for $\mu> 0.35$. Remarkable enough the texture of Poincar\'e map is qualitatively the same and similar to the one of Duffing model, for all values of the deformability parameter for which chaotic atttractors emerge. This suggests that the hyperbolic double-well map proposed provides an ideal generalization of the Duffing model by enabling full parametric control of the systerm dynamics.
\section{\label{sec5} Concluding remarks}
Damped and driven oscillators moving in the field of force created by a double-well potential abound in nature. They are observed in mechanical engineering, as for instance in rotating machinery, radial loads in bearing rotor systems, or gravity loads in cracked rotor systems. Take for instance a slender steel fixed on a bracket of a rigid insulating frame, with the bottom of the slender steel placed in between two permanent magnets and the rigid frame driven by a periodic excitation. This mechanical setup can be described by the so-called Holmes-Duffing model, and has been the subjsect of a great deal of attention (see e.g. \cite{dh1,dh2}). In electrical engineering, bistable circuits can be designed where bistability is produced by a nonlinear feedback component such as a VARACTOR, a RESISTOR or a MEMRISTOR \cite{d1,d2,d3} whcih are two-terminal nonlinear electrical components connecting electric charge or currents and capacitors, resistances or magnetic flux. In VARACTOR diodes a bistable feedback can be caused by a CV characteristics that favors hysteresis processes in driven active RLC circuits \cite{hd3,hd4}. For these physical systems the Duffing model has been a rich paradigm, however the great variety of existing bistable systems motivates the quest for a more physically realistic theoretical description of their dynamical properties. To this last point the bistable feature of the Duffing model is related to the so-called $\phi^4$ or Ginzburg-Landau potential, which actually is a quartic polynomial and therefore can readily be looked out as a truncated expansion of a non-polynomial two-minima function. In this study we considered one such realistic double-well potential, it is an hyperbolic function of the displacement and has the virtue of enabling the account of smooth changes in positions of the two symmetric potential wells without affecting the barrier height. In the example of mechanical system consisting of steel slender mentioned above, the change in positions of the potential minima would correspond to a continuous shift of positions of the two magnets. In the context of active electrical circuits, a nonlinear feedback governed by an hyperbolic dependence either of the capacitance with voltage or of the resistance with current will promote hysteresis phenomena as established by Leon Chua in several study contexts \cite{a24,d1,d5,d6}. \\
In summary we have proposed a model of damped and driven oscillator, moving in the field of force created by a double-well potential. Contrary to the standard Duffing model for which the double-well potential is the $\phi^4$ field, the double-well substrate proposed is non-polynomial and more precisely an hyperbolic double-well potential. It has the advantage that positions of its two stable equilibrium states can be shifted by varying a deformability parameter, a feature that can be determinant in the study of systems whose structures are flexibles and hence their characteristic points are not fix during their motions. Moreover for a specific value of the deformability parameter, the hyperbolic double-well oscillator model reduces exactly to the standard Duffing model. The fact that most of the physical attributes of the standard Duffing oscillator, such as the existence of period-doubling transitions, limit cycles and chaotic orbits and so on, are also observed in the proposed model, makes the model an ideal generalization of the Duffing model.
\begin{acknowledgments}
Work supported by the Alexander von Humboldt foundation.
\end{acknowledgments}
\section*{Author declarations}
\subsection*{Conflict of Interest}
The author has no conflict of interest to disclose.
\subsection*{Data availability}
The data that support the findings of this study are available within the article.



\begin{thebibliography}{}\label{sec:references}

\bibitem{a1}J. A. Krumhansl and J. R. Schrieffer, "Dynamics and statistical mechanics of a one-dimensional model Hamiltonian for structural phase transitions", {\it Phys. Rev.} B \textbf{11}, 3535-3545 (1975).
\bibitem{a2}J. F. Currie, J. A. Krumhansl, A. R. Bishop and S. E. Trullinger, "Statistical mechanics of one-dimensional solitary-wave-bearing scalar fields: Exact results and ideal-gas phenomenology", {\it Phys. Rev.} B \textbf{22}, 477-496 (1980).
\bibitem{a3}L. D. Landau and E. M. Lifshitz, 1958 {\it Statistical Physics} (Pergarmon, London).
\bibitem{a4}A. M. Dikand\'e and T. C. Kofan\'e, "Oscillatory motions of solitons in finite inhomogeneous structures", {\it J. Phys. Condens. Matt.} {\bf 6}, 6229-6236 (1994).
\bibitem{a5}H. Roohi and S. Bagheri, "Influence of substitution on the strength and nature of CH$\cdots$N hydrogen bond in XCCH$\cdots$NH3 complexes", {\it Int. J. Quantum Chem.} {\bf 111}, 961-969 (2011).
\bibitem{a51}J. S. Plant and H. H. Greenwood, "J. S. Plant and H. H. Greenwood", {\it Int. J. Quantum Chem.} {\bf 35}, 385-393 (1989).
\bibitem{a6}H. H. Greenwood and J. S. Plant, "Ab initio theory of hydrogen bonding in vitamin B6", {\it Int. J. Quantum Chem.} {\bf 30}, 127-136 (1986).
\bibitem{a7}Y. Zhang, Z. Cao, J. Z. Zhang and F. Xia, "Double-well wltra-coarse-grained model to describe protein conformational transitions", {\it J. Chem. Theory Comput.} {\bf 16}, 6678 (2020).
\bibitem{a8}G. Williams and A. Toon, "Protein folding pathways and state transitions described by classical equations of motion of an elastic network model", {\it Protein Science} {\bf 19}, 2451-2461 (2010).
\bibitem{a81}J. Roth, M. D. Koch and A. Rohrbach, "Dynamics of a Protein Chain Motor Driving Helical Bacteria under Stress", {\it Biophys. J.} {\bf 114}, 1955 (2018).
\bibitem{a9}N. Mukherjee and A. K. Roy, "Quantum confinement in an asymmetric double‐well potential through energy analysis and information entropic measure", {\it Annal. der Phys.} {\bf 528}, 412 (2016).
\bibitem{a10}L. Wang, Q. Zhang, F. Xu, X.-D. Cui and Y. Zheng, {\it Int. J. Quantum Chem.} {\bf 115}, 208 (2015).
\bibitem{a12}C. Wagner and Th. Kiefhaber, "Intermediates can accelerate protein folding", {\it PNAS} {\bf 96}, 6716-6721 (1999).
\bibitem{a13}K. L. Sebastian and A. Debnath, "Polymer in a double well: dynamics of translocation of short chains over a barrier", {\it J. Phys.: Condens. Matter} {\bf 18}, S283-S296 (2006).
\bibitem{a14}S. Lee and W. Sung, "Coil-to-stretch transition, kink formation, and efficient barrier crossing of a flexible chain", {\it Phys. Rev.} E {\bf 63}, 021115 (2001).
\bibitem{a15} I. Kovacic and M. J. Brennan: {\it The Duffing equation: Nonlinear oscillators and their behaviour} (John Wiley and Sons, 2011).
\bibitem{a16}H. Konwent, P. Machnikowski, P. Magnuszewski and A. Radosz, "Some properties of double-Morse potentials", {\it J. Phys.} A{\bf 31}, 7541 (1998).
\bibitem{a17}H. Konwent, "One-dimensional Schr\"dinger equation with a new type double-well potential", {\it Phys. Lett.} A {\bf 118}, 467 (1986).
\bibitem{a18}M. Razavy, "Eigenvalues of the Schr\"odinger Equation for a Periodic Potential", {\it Am. J. Phys.} {\bf 48}, 285 (1980).
\bibitem{a19}C. A. Downing, "On a solution of the Schr\"odinger equation with a hyperbolic double-well potential ", {\it J. Math. Phys.} {\bf 54}, 072101 (2013).
\bibitem{a20}R. R. Hartmann, "Bound states in a hyperbolic asymmetric double-well ", {\it J. Math. Phys.} {\bf 55}, 012105 (2014).
\bibitem{a21}A. M. Dikand\'e and T. C. Kofan\'e, "Class of deformable double-well potentials with exact kink solutions", {\it Solid State Commun.} {\bf 89}, 559 (1994).
\bibitem{a22}A. M. Dikand\'e and T. C. Kofan\'e, "Exact kink solutions in a new non-linear hyperbolic double-well potential", {\it J. Phys. Condens. Matter.} {\bf 3}, L5203 (1991).
\bibitem{a23}A. M. Dikand\'e and T. C. Kofan\'e, "Phonons response to nonlinear excitations in a new parametrized double-well one-site potential lattice", {\it Solid State Commun.} {\bf 86}, 749 (1993).
\bibitem{a24} L. Chua, {\it Visions of Nonlinear Science in the 21st Century}, in {\it Nonlinear
Science Series} (vol. 26, World Scientific, Singapore 1999).
\bibitem{luth}H. A. Luther, "An Explicit Sixth-Order Runge-Kutta Formula", {\it Math. Comp.} {\bf 22}, 434 (1968).
\bibitem{abra}M. Abramovich and I. Stegun, {\it Handbook of Mathematical Functions} (Dover, New York, 1964).
\bibitem{solduf1}J. Tezcan and J. K. Hsiao, "Periodic solutions of the Duffing equation", {\it Struc. Eng. Mechan.} {\bf 30}, 593 (2008).
\bibitem{solduf2}L. Burra and F. Zanolin, {\it The Duffing Equation: Periodic solutions and chaotic dynamics} (Infosys Science Foundation Series in Mathematical Sciences, Springer, 2025)
\bibitem{solduf3}H. Wang and Y. Li, "Periodic solutions for Duffing equations", {\it Nonl. Anal. Methods and Appli.} {\bf 24}, 961 (1995).
\bibitem{solduf4}A. H. Salas, J. E. C. Hernandez and L. J. M. Hernandez, "The Duffing Oscillator Equation and Its Applications in Physics", {\it Mathematical Problems in Engineering} (Wiley, 2021).
\bibitem{wig}S. Wiggins, {\it Introduction to Applied Nonlinear Dynamical Systerms and Chaos} (springer-Verlag, New York, 1990).
\bibitem{chao}H. J. Korsch and H. J. Jodl, {\it Chaos: A Program collection for the PC} ($2^{nd}$ ed., Springer-Verlag Berlin Heidelberg, 1998.)
\bibitem{toguy}J. Kengne, J. C. Chedjou, G. kenne, K. Kyamakya and G. H. Kom, "Analog circuit implementation and synchronization of a system consisting of a van der Pol oscillator linearly coupled to a Duffing oscillator", {\it Nonlinear Dyn.} \textbf{70}, 2163-2173 (2012).
\bibitem{dh1}M. Wang, L. Zhou, E. Chen and P. Liu, "Dynamical characterization of a Duffing–Holmes system containing nonlinear damping under constant excitation", {\it Chaos, Solitons and Fractals} \textbf{175-1}, 113926 (2023).
\bibitem{dh2}T. Chen, X. Cao and D. Niu, "Model modification and feature study of Duffing oscillator", {\it J. Low. Freq. Noise, Vib. and Active Cont.} \textbf{4}, 230 (2021).
\bibitem{d1}F. Z. Wang, L. Li, L. Shi, H. Wu and L. O. Chua, "$\Phi$ memeristor: Real memristor found", {\it J. Appl. Phys.}
\textbf{125}, 054504 (2019).
\bibitem{d2}M. Chen, W. Xue, X. Luo, Y. Zhang and H. Wu, "Effects of coupling memristors on synchronization of two identical memristive Chua's systems", {\it Chaos, Solitons and Fractals} \textbf{174}, 113780 (2023).
\bibitem{d3}N. A. Khan, M. A. Qureshi and N. A. Khan, "Evolving Tangent hyperbolic memristor based 6D chaotic model with fractional order derivative: analysis and applications", {\it Par. Diff. Eq. Appl. Math.} \textbf{7}, 100505 (2023).
\bibitem{hd3}B. K. Jones and G. Trefan, "The Duffing oscillator: A precise electronic analog chaos demonstrator for the undergraduate laboratory", {\it Amer. J. Phys.} \textbf{69}, 464-469 (2001).
\bibitem{hd4}P. K. Shaw, M. S. Janaki, A. N. S. Iyengar, T. Singla and P. Parmananda, "Antiperiodic oscillations in a forced Duffing oscillator", {\it Chaos, Solitons and Fractals} \textbf{78}, 256 (2015).
\bibitem{d5}L. Chua, "Memristor-The missing circuit element", {\it IEEE Trans.Circ. Theory CT} \textbf{18}, 507 (1971).
\bibitem{d6}L. Chua, "Resistance switching memories are memristors", {\it Appl. phys.} A\textbf{102}, 765(2011).
\end{thebibliography}
\end{document}